# Construction and Optimization of TRNG Based Substitution Boxes for Block Encryption Algorithms

**Muhammad Fahad Khan [1, 2, *], Khalid Saleem [1], Mohammed Alotaibi [3], Mohammad Mazyad Hazzazi [4], Eid Rehman [2], Aaqif Afzaal Abbasi [2], Muhammad Asif Gondal [5]**

[1] Quaid e Azam University, Islamabad, 45320, Pakistan
[2] Foundation University Islamabad, Islamabad, 44000, Pakistan
[3] College of Business Administration, University of Tabuk, Tabuk, 71491, Saudi Arabia
[4] Department of Mathematics, College of Science, King Khalid University, Abha, 61413, Saudi Arabia
[5] Department of Mathematics and Sciences, Dhofar University, Salalah, 211, Oman
[*]Corresponding Author: M.Fahad Khan. mfkhan@cs.qau.edu.pk
Received: XX Month 202X; Accepted: XX Month 202X

**Abstract**: Internet of Things is an ecosystem of interconnected devices that are accessible through the internet. The recent research focuses on adding more smartness and intelligence to these edge devices. This makes them susceptible to various kinds of security threats. These edge devices rely on cryptographic techniques to encrypt the pre-processed data collected from the sensors deployed in the field. In this regard, block cipher has been one of the most reliable options through which data security is accomplished. The strength of block encryption algorithms against different attacks is dependent on its nonlinear primitive which is called Substitution Boxes. For the design of S-boxes mainly algebraic and chaos-based techniques are used but researchers also found various weaknesses in these techniques. On the other side, literature endorse the true random numbers for information security due to the reason that, true random numbers are purely non-deterministic. In this paper firstly a natural dynamical phenomenon is utilized for the generation of true random numbers based S-boxes. Secondly, a systematic literature review was conducted to know which metaheuristic optimization technique is highly adopted in the current decade for the optimization of S-boxes. Based on the outcome of Systematic Literature Review (SLR), genetic algorithm is chosen for the optimization of s-boxes. The results of our method validate that the proposed dynamic S-boxes are effective for the block ciphers. Moreover, our results showed that the proposed substitution boxes achieve better cryptographic strength as compared with state-of-the-art techniques.

**Keywords:** IoT security; Sensors data encryption; Substitution box generation; True Random Number Generators (TRNG); Heuristic optimization; Genetic algorithm;

## 1 Introduction

IoT created a new paradime by connecting various devices and everyday objects to the network but now days IoT devices have facing serious security issues to protect the confidentiality of transmitted data. Block ciphers are most reliable option for ensuring the data confidentiality. The strength of block ciphers against different attacks depends on their substitution boxes which provides confusion in the cryptosystem. S-box is a mapping of n-bits inputs to m-bits output and it can be seen as a Boolean function and this can be viewed as $F: F_2^n \rightarrow F_2^m$. In the literature mainly algebraic methods and chaotic maps are used for the construction of S-boxes although there methods have favorable characteristics for S-box designs but many researchers pointed weaknesses in these methods such as Short Quantity of randomness [1-6], Interpolation attacks [7-10], eXtended Sparse Linearization (XSL) attack [11-15], Initial state limitation [16-17], Frail chaos [18-19], Small number of input parameters [20-23], Dynamical degradation [24-27], Gröbner basis attack [27-29], eXtended Linearization(XL) attacks [30-37].

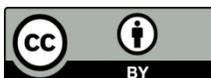

On the other side, randomness is a fundamental aspect of many processes in nature and an indubitably valuable resource for cryptography. Researchers endorse the true random numbers in cryptography due to the fact that true random numbers are irreversible, unpredictable, and unreproducible even if their internal structure and response history is known to the adversaries. We are using the lightning strike phenomenon as an entropy source to generate true random numbers. In this research first of all, we intake the locations of lightning strikes in the row data format from the standard repository of National Aeronautics and Space Administration(NASA) which name is Lightning Detection and Ranging System (LDAR) [38]. Secondly, a novel technique is proposed for the generation of TRNG based substitution boxes. Thirdly genetic algorithm applied for the optimization of newly generated S-boxes but before this phase, a systematic literature review was conducted to know which metaheuristic optimization technique is highly adopted in the last 10 years. Based on the outcome of SLR, a genetic algorithm is chosen. The remaining paper is structured as follows; sections 2 and 3 present our contribution and the proposed methodology respectively, section 4 explains the results and evaluation section, section 5 shows the conclusion section.

## 2 Contribution

The main contributions of this research are follows:
- A novel method is proposed for the generation of substitution boxes of symmetric encryption algorithms based on true random numbers.
- Multi population genetic algorithm is applied for optimization of substitution boxes.
- A systematic literature review was conducted to know which metaheuristic optimization technique is highly adopted in the last 10 years for the optimization of cryptographic substitution boxes.
- A novel technique proposed for the true random numbers generation.

## 3 Proposed Design Methodology

The proposed technique has three phase's first phase is true random bits extraction, second phase is construction of substitution boxes and the third phase is optimization of substitution boxes. These three phases are explained in the following and the proposed system architecture is depicted in Fig. 1.

### 3.1 True Random Bits Extraction

To calculate the difference between each location of the lightning strike with other lightning strike locations, firstly we acquire the lightning strike locations (in the row data format) from the standard repository of NASA [38]. Lightning detection and the ranging system is a volumetric lightning mapping system, which stores the real-time location of the striking point. The format of the row data is dd,hh,mm,ss,ll,xx,yy, zz, where dd represents the day of the month, hh represents the hour, mm represents the minute, ss represents second, ll represents microsecond. xx and yy represent the distance in meters from site-1 to the east direction and north direction respectively where zz represents the distance in meters above the surface of the earth. Secondly, each location of the 1st lightning strike is subtracted from the other locations of the lightning strikes. For the experiment, data of the randomly picked 9613227 lightning strikes are processed from the proposed algorithm and then resultant binary stream enhanced through the Von Neumann extractor, after this, these random bits are tested through the National Institute of Standards and Technology (NIST) statistical test suite. Results of the NIST statistical test (shown in Appendix A) proved that lightning strikes are useful entropy source for the generation of true random numbers.

### 3.2 Construction of Substitution Boxes

In the first step of dynamic substitution boxes construction, an algorithm proposed which name is

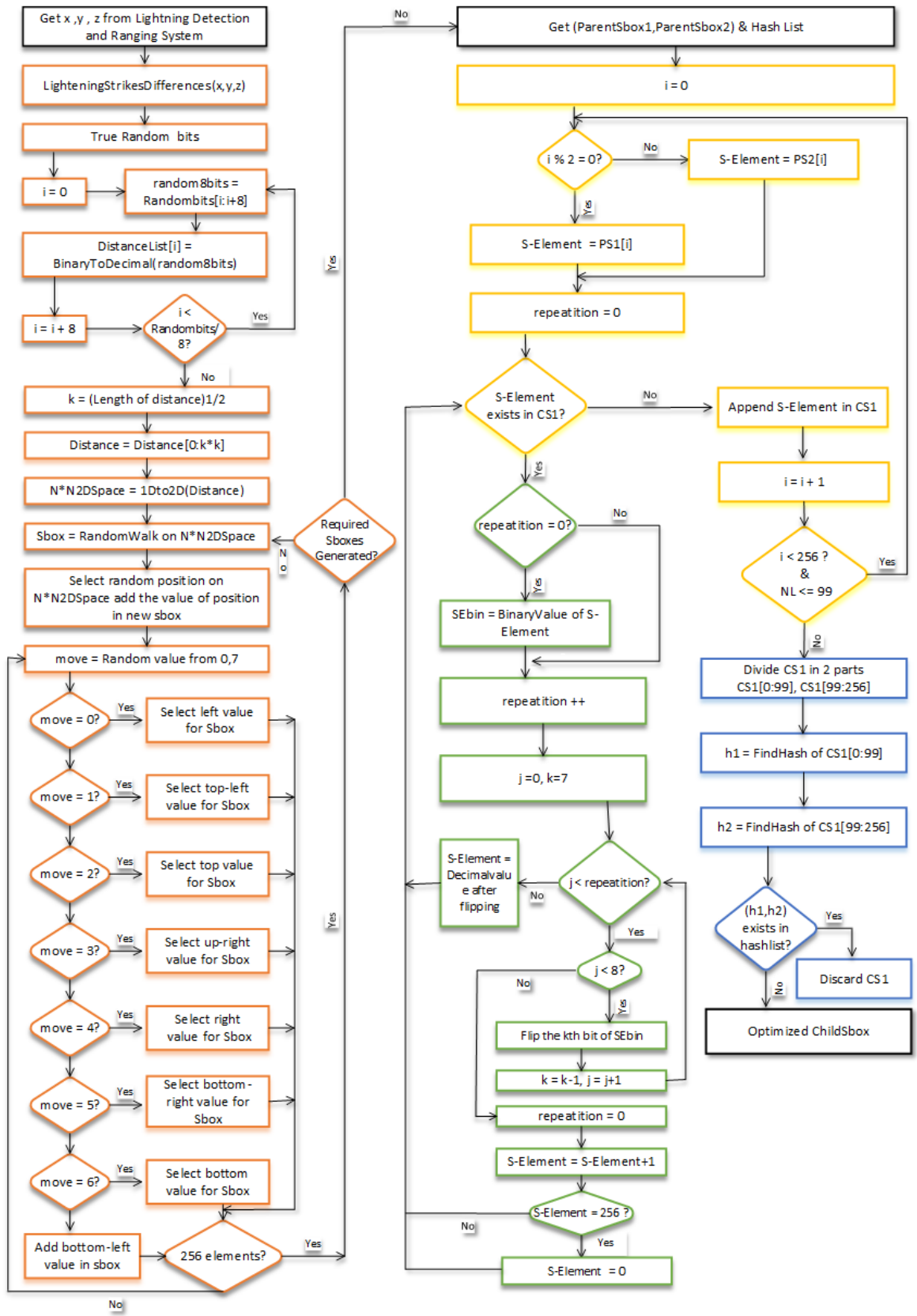

**Figure 1:** Proposed System Architecture

| **Algorithm:** SboxConstruction( RandomBits, totalSbox ) |
|---|
| **Input:** RandomBits string; integer totalSbox |
| **Output:** 1D array of integer values Sbox[ ] |
| 1: Distance ← list( ) |
| 2: z ← 0 |
| 3: **for** i ← 0 …( | RandomBits | **div** 8 ) **do** |
| 4:     random8bits ← substring( RandomBits,z,z+8 ) |
| 5:     Distance.append( BinaryToDecimal( random8bits) ) |
| 6:     z ← z + 8 |
| 7: **end for** |
| 8: k ← ( | Distance | ) $^{1/2}$ |
| 9: Distance ← Distance[ 0 : k $^2$ ] |
| 10:     NxN2DSpace ← 1DTo2DArray( Distance, ( k,k ) ) |
| 11: n ← 0 |
| 12: **while** ( n != totalSbox)  **do** |
| 13:     tempSbox ← RandomWalk( NxN2DSpace, k) |
| 14:     Sbox.append(tempSbox) |
| 15:     n++ |
| 16: **end while** |
| 17: **return** Sbox |

SboxConstruction(). This algorithm takes, true random bits from the last phase and a total number of required S-boxes from the user as parameters. In the second step SboxConstruction() algorithm, further calls the RandomWalk() algorithm by passing the two-dimensional N * N space of truly random bits. On every run RandomWalk() algorithm generates a single dynamic s-box by using the 8 states random walk rule ( left = 0, left up = 1, up word = 2, right up = 3, right = 4, right down = 5, down = 6, left down = 7). SboxConstruction() algorithm is presented in the above. Randomly we plotted the two RandomWalk() runs as sample in the Fig. 2a,2b and their resultant S-boxes in the Tab. 1a and 1b. For the results, the total number of ten thousand S-boxes are constructed, where the nonlinearity score of 1674 S-boxes are ≤ 99, nonlinearity score of 3277 S-boxes are between 100 to 102, nonlinearity score of 4126 S-boxes are between 103 to 104, nonlinearity score of 923 S-boxes are between 105 to 106. The nonlinearity score of our sample S-boxes 2a, 2b is 105 and 104 respectively. As sample TRNG based S-box is shown in the Appendix B.

**Table 1a:** Proposed Sbox-1

| 155 | 253 | 60 | 189 | 42 | 238 | 93 | 209 | 202 | 129 | 164 | 75 | 240 | 85 | 151 | 254 |
|---|---|---|---|---|---|---|---|---|---|---|---|---|---|---|---|
| 44 | 138 | 251 | 28 | 223 | 230 | 95 | 62 | 231 | 121 | 149 | 31 | 180 | 57 | 161 | 73 |
| 203 | 72 | 84 | 227 | 218 | 210 | 56 | 243 | 245 | 190 | 135 | 9 | 69 | 255 | 219 | 166 |
| 140 | 198 | 239 | 225 | 250 | 143 | 120 | 63 | 134 | 103 | 186 | 207 | 2 | 64 | 27 | 144 |
| 241 | 37 | 108 | 92 | 98 | 65 | 220 | 79 | 4 | 249 | 228 | 24 | 205 | 10 | 102 | 156 |
| 8 | 122 | 119 | 185 | 188 | 58 | 236 | 216 | 226 | 17 | 217 | 13 | 35 | 237 | 78 | 242 |
| 80 | 33 | 127 | 125 | 14 | 83 | 192 | 66 | 221 | 169 | 48 | 112 | 59 | 100 | 146 | 74 |
| 199 | 97 | 141 | 52 | 173 | 107 | 224 | 68 | 55 | 152 | 247 | 110 | 157 | 181 | 170 | 196 |
| 131 | 32 | 104 | 86 | 96 | 128 | 50 | 208 | 158 | 244 | 214 | 159 | 234 | 142 | 22 | 118 |
| 81 | 43 | 6 | 89 | 187 | 7 | 177 | 67 | 204 | 171 | 5 | 235 | 195 | 123 | 0 | 139 |
| 109 | 3 | 178 | 172 | 18 | 54 | 76 | 53 | 99 | 88 | 212 | 46 | 21 | 39 | 15 | 201 |
| 101 | 87 | 246 | 184 | 165 | 233 | 105 | 179 | 232 | 206 | 49 | 70 | 252 | 126 | 191 | 147 |
| 114 | 71 | 229 | 26 | 1 | 29 | 132 | 248 | 91 | 154 | 25 | 211 | 16 | 106 | 34 | 111 |
| 116 | 183 | 168 | 117 | 193 | 51 | 47 | 137 | 115 | 200 | 174 | 153 | 133 | 20 | 213 | 194 |
| 45 | 40 | 124 | 150 | 130 | 11 | 77 | 175 | 12 | 222 | 19 | 160 | 30 | 182 | 82 | 41 |
| 61 | 113 | 36 | 167 | 38 | 136 | 145 | 163 | 90 | 94 | 162 | 23 | 215 | 176 | 197 | 148 |

**Table 1b:** Proposed Sbox-2

| | | | | | | | | | | | | | | | |
|---|---|---|---|---|---|---|---|---|---|---|---|---|---|---|---|
| 211 | 238 | 81 | 126 | 89 | 185 | 244 | 26 | 139 | 143 | 23 | 73 | 209 | 102 | 132 | 120 |
| 34 | 174 | 78 | 159 | 165 | 114 | 138 | 181 | 10 | 28 | 221 | 250 | 44 | 189 | 110 | 191 |
| 155 | 128 | 223 | 75 | 164 | 72 | 214 | 108 | 25 | 245 | 201 | 213 | 2 | 109 | 178 | 144 |
| 216 | 96 | 123 | 196 | 197 | 121 | 150 | 86 | 35 | 169 | 27 | 90 | 166 | 163 | 56 | 158 |
| 117 | 147 | 15 | 6 | 47 | 17 | 100 | 252 | 57 | 79 | 156 | 212 | 198 | 218 | 131 | 162 |
| 101 | 186 | 30 | 104 | 242 | 202 | 146 | 122 | 133 | 253 | 103 | 130 | 87 | 16 | 192 | 168 |
| 54 | 205 | 125 | 149 | 33 | 107 | 153 | 32 | 129 | 179 | 12 | 173 | 251 | 157 | 195 | 42 |
| 224 | 227 | 20 | 207 | 180 | 36 | 145 | 84 | 231 | 200 | 255 | 61 | 97 | 43 | 170 | 91 |
| 11 | 49 | 210 | 148 | 204 | 24 | 13 | 215 | 52 | 46 | 183 | 193 | 171 | 154 | 222 | 88 |
| 137 | 94 | 85 | 40 | 167 | 95 | 140 | 45 | 106 | 237 | 99 | 74 | 234 | 175 | 21 | 151 |
| 4 | 119 | 77 | 184 | 105 | 92 | 83 | 243 | 177 | 226 | 51 | 134 | 39 | 182 | 68 | 236 |
| 82 | 118 | 248 | 0 | 70 | 247 | 55 | 67 | 37 | 8 | 230 | 136 | 48 | 208 | 188 | 142 |
| 41 | 232 | 246 | 235 | 71 | 113 | 127 | 69 | 5 | 116 | 22 | 249 | 64 | 80 | 93 | 50 |
| 53 | 219 | 1 | 161 | 76 | 160 | 172 | 176 | 225 | 63 | 152 | 254 | 187 | 115 | 98 | 239 |
| 31 | 29 | 62 | 220 | 18 | 112 | 206 | 194 | 229 | 241 | 240 | 217 | 228 | 111 | 141 | 38 |
| 135 | 203 | 3 | 233 | 19 | 65 | 14 | 190 | 199 | 58 | 66 | 59 | 60 | 9 | 124 | 7 |

**Table 1c:** Optimized Sbox-3

| | | | | | | | | | | | | | | | |
|---|---|---|---|---|---|---|---|---|---|---|---|---|---|---|---|
| 162 | 5 | 201 | 199 | 126 | 220 | 204 | 9 | 155 | 242 | 2 | 39 | 166 | 221 | 133 | 15 |
| 63 | 192 | 73 | 164 | 254 | 23 | 11 | 185 | 247 | 34 | 98 | 149 | 40 | 22 | 96 | 33 |
| 176 | 75 | 113 | 53 | 50 | 165 | 44 | 29 | 132 | 156 | 169 | 206 | 225 | 85 | 0 | 208 |
| 36 | 103 | 64 | 218 | 198 | 122 | 137 | 136 | 104 | 159 | 118 | 88 | 154 | 32 | 193 | 57 |
| 183 | 135 | 160 | 24 | 228 | 152 | 234 | 172 | 17 | 177 | 232 | 70 | 175 | 210 | 14 | 31 |
| 142 | 188 | 227 | 141 | 151 | 76 | 219 | 131 | 119 | 197 | 212 | 251 | 65 | 100 | 97 | 145 |
| 190 | 68 | 158 | 110 | 214 | 61 | 146 | 116 | 3 | 238 | 101 | 49 | 181 | 54 | 239 | 92 |
| 179 | 147 | 30 | 143 | 109 | 170 | 223 | 215 | 157 | 51 | 250 | 106 | 153 | 13 | 241 | 79 |
| 129 | 140 | 78 | 18 | 248 | 6 | 196 | 252 | 58 | 89 | 184 | 134 | 115 | 59 | 203 | 124 |
| 81 | 112 | 195 | 84 | 74 | 235 | 82 | 28 | 105 | 243 | 237 | 230 | 95 | 173 | 80 | 94 |
| 138 | 83 | 66 | 12 | 38 | 45 | 87 | 10 | 191 | 121 | 86 | 99 | 189 | 217 | 249 | 168 |
| 236 | 102 | 231 | 1 | 205 | 107 | 62 | 202 | 233 | 253 | 27 | 56 | 144 | 72 | 8 | 125 |
| 167 | 55 | 128 | 25 | 255 | 229 | 187 | 178 | 35 | 246 | 171 | 226 | 26 | 240 | 47 | 16 |
| 211 | 77 | 117 | 41 | 20 | 67 | 123 | 180 | 111 | 91 | 174 | 224 | 7 | 46 | 52 | 127 |
| 120 | 108 | 182 | 244 | 222 | 207 | 150 | 60 | 130 | 148 | 42 | 139 | 194 | 93 | 21 | 216 |
| 209 | 161 | 48 | 245 | 4 | 37 | 19 | 43 | 114 | 90 | 69 | 213 | 186 | 71 | 163 | 200 |

### 3.3 Optimization of Substitution Box

The most important strength of true random-based construction of S-boxes is that, these S-boxes are immune against the attacks which are mentioned in the introduction section. This is our first goal which is achieved in the above sections 3.2 by using the true randomness. The 2nd goal of this research is to find out, which metaheuristic optimization technique is effective (highly adopted) in existing literature for the optimization of S-boxes (which are not based on true randomness). To discover the answer to this question, a SLR was conducted over the last 10 years. Query wise search results of SLR is mentioned in the Tab. 2 and based on the SLR recommendation, a technique is presented in the following for the optimization of newly constructed S-boxes using the genetic algorithm. Reverse S-box algorithm for the TRNG and GA based S-boxes is shown in the Appendix C.

**Table 2:** Query wise search results

| Query | IEEE | Springer | ACM | Elsevier |
|---|---|---|---|---|
| Genetic Algorithm and S-boxes | 14 | 241 | 33 | 62 |
| Genetic Algorithm and Substitution box | 5 | 93 | 24 | 29 |
| Genetic Algorithm and Nonlinear Block Cipher Primitive | 4 | 0 | 66 | 0 |
| Genetic Algorithm and Confusion component | 5 | 3 | 54 | 0 |
| Simulate Annealing and S-boxes | 4 | 99 | 13 | 19 |
| Simulate Annealing and Substitution box | 1 | 39 | 9 | 6 |
| Simulate Annealing and Nonlinear Block Cipher Primitive | 0 | 0 | 14 | 0 |
| Simulate Annealing and confusion component | 0 | 0 | 17 | 0 |
| Tabu Search and Substitution box | 0 | 4 | 10 | 2 |
| Tabu Search and S-boxes | 2 | 9 | 12 | 8 |
| Tabu Search and Nonlinear Block Cipher Primitive | 0 | 8 | 14 | 0 |
| Tabu Search and Confusion component | 0 | 0 | 14 | 0 |
| Particle Swarm Optimization and S-boxes | 2 | 79 | 51 | 21 |
| Particle Swarm Optimization and Substitution box | 1 | 30 | 48 | 8 |
| Particle Swarm Optimization and Nonlinear Block Cipher Primitive | 0 | 0 | 48 | 0 |
| Particle Swarm Optimization and Confusion component | 5 | 0 | 39 | 0 |
| Artificial Bee colony Optimization and Substitution box | 1 | 9 | 48 | 1 |
| Artificial Bee colony Optimization and S-boxes | 2 | 8 | 50 | 3 |
| Artificial Bee colony Optimization and Nonlinear Block Cipher Primitive | 0 | 0 | 1 | 0 |
| Artificial Bee colony Optimization and Confusion component | 0 | 0 | 6 | 0 |

GA is based on the Charles Darwin's theory of natural selection and survival of the fittest In the standard genetic algorithm, selection, crossover, and mutation are the common processes. In our problem for the selection process, all the substitution boxes whose nonlinearity score lies in the range of 100 to 106 are acquired as the initial population. For the crossover process, we used the one-point crossover strategy in which pair of parents exchange the half part of each parent to each other and generate new offspring. Complete crossover process is represented with yellow color in the flowchart which is depicted in Fig. 1. Widely in the literature [39-44], the nonlinearity score is the highly desirable property that's why we also chose the nonlinearity score as the fitness value. In the conventional genetic algorithm, crossover and mutation are the two independent processes but here for the substitution box generation, we combine the crossover and mutation process together. The reason is that, the substitution boxes generated by the crossover operation usually do not satisfy the bijective property due to repeated elements. So, repetition must be removed from the new offspring to achieve the bijective property and for that purpose, a simple strategy is adopted in the mutation process. In the mutation process we flipped the one bit from right side to the left side, and after flipping each bit, the corresponding number is checked within the s-box. If the corresponding number is unique, then we stop the flipping process, otherwise we flip the next bit and so on. In the numerous cases, we achieved the bijective property but in the few cases after flipping the eight bits, we did not get the unique element of the s-box and in those cases, we simply add one to the original element from where we start the flipping process. The mutation process is represented with green color in the flowchart of Fig. 1. From the results, we observed that numerous S-boxes and their subsets are repeated therefore we used SHA-3 based searching for removal and this process is shown with blue color in the flowchart. Here, for the experiment hundred thousand S-boxes are optimized and to examine the performance, we compared the highest quality ten thousand optimized S-boxes with the ten thousand S-boxes which are already constructed through the TRNG based technique. After optimization, it can be clearly seen from Fig. 3 that GA improves the overall quality of the S-Boxes. After optimization process, there are no S-box having non-linearity of less than and equal to 99 where TRNG based S-box construction technique 1674 S-boxes in that range. Similarly, GA produce more S-boxes in the range of 105 and 106 as compared to TRNG construction method. Furthermore, GA produce 616 new S-boxes having non-linearity of 107 and 108 and these S-Boxes were not discovered in TRNG based approach.

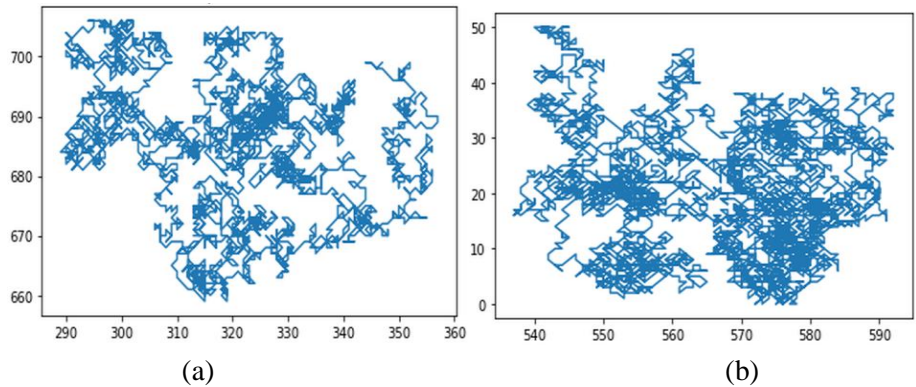

(a) (b)

**Figure 2:** Plotted S-boxes (a) 8-steps random walk of Sbox-1 (b) 8-steps random walk of Sbox-2

## 4. Results and Evaluation

In this section sample S-boxes of section 3.2 are evaluated through the S-box evaluation criteria which are shown in the following 4.1 to 4.5.

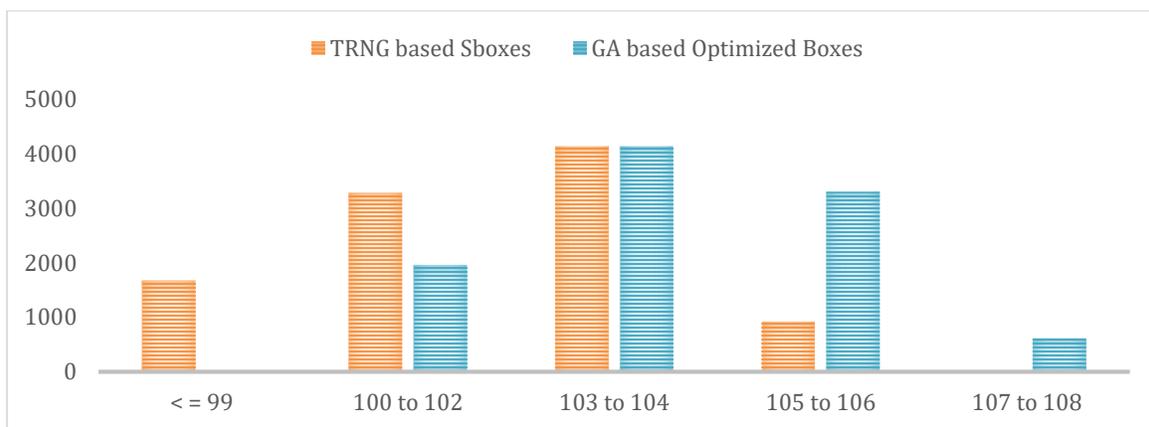

**Figure 3:** Nonlinearity score of TRNG and GA based S-box construction

### *4.1 Nonlinearity*

Out of all cryptographic properties, nonlinearity is said to be the most significant. For a strong encryption scheme, the mapping between input and output in an S-box must be nonlinear. Nonlinearity can be defined as the smallest distance of Boolean function to the set of affine functions. To get the closet affine function in the Boolean truth table, the total number of bits altered needs to be determined by the nonlinearity score. In the above Fig. 3 we saw that six hundred sixteen S-boxes attains a 108 nonlinearity score and from these S-boxes, we picked randomly one S-box as a sample which is shown in Tab. 1c. The Nonlinearity scores of proposed S-boxes are better or equal to the state-of-the-art. S-boxes which are shown in Tab. 3. By using Walsh spectrum, the nonlinearity of Boolean function is determined as:

$$N_g = 2^{n-1}(1 - 2^{-n} \max_{\varphi \in GF(2^n)} |S(g)(\varphi)|) \quad (1)$$

Where $S_{(g)}(\varphi)$ is defined as: $S_{(g)}(\varphi) = \sum_{\varphi \in GF(2^n)} (-1)^{g(x) \oplus x.\varphi}$ (2)

Where $\varphi$ is a n-bit vector and $\varphi \in GF(2^n)$. The dot product between x and $\varphi$ is denoted by $x.\varphi$

$x.\varphi = x1 \oplus \varphi1 + x2 \oplus \varphi2 \cdot + xn \oplus \varphi n$.

**Table 3:** Nonlinearity of various S-boxes

| S-box | Nonlinearity | S-box | Nonlinearity |
|---|---|---|---|
| Ozkaynak [45], 2020 | 104 | EI-Latif [46], 2020 | 107 |
| Zahir [47], 2020 | 104 | EI-Latif [48], 2020 | 106 |
| Bin [49], 2020 | 104 | Islam [50], 2017 | 108 |
| Belazi [51], 2017 | 108 | Ozkaynak [52], 2018 | 108 |
| cavusoolu [53], 2017 | 106 | Solami [54] 2018 | 108 |
| Lambi [55], 2017 | 108 | Lambić [56], 2020 | 106 |

*4.2 Strict Avalanche Criteria (SAC)*

SAC computes the number of output bits altered caused by inverting a single bit of input. To make the system more reliable, we need to deviate output vector with half probability when one input bit is inverted. To evaluate the SAC property, dependency matrix are used. For an S-box that can satisfy SAC property, all values need to be close to ideal value of 0.5 in its dependence matrix. The SAC value of our randomly picked optimized S-box is 0.501465 which satisfies the avalanche criteria. The offsets of the dependence matrix can be determined by:

$$S(g) = \frac{1}{n^2} \sum_{1 \le r \le n} \sum_{1 \le w \le n} |\frac{1}{2} - Q_{r,w}(g)| \quad (3)$$

Where $Q_{r,w}(g) = 2^{-n} \sum_{x \in B^n} gw(x) \oplus gw(x \oplus e_r)$ (4)

$e_r = [\theta r, 1 \; \theta r, 2 \ldots \theta r, n]^T \quad \theta_{r,w} = 0, r \ne w \quad \text{or} \quad \theta r, w = 1, r = w$

*4.3 BIT Independent Criterion (BIC)*

BIC is another cryptographic metric used to measure the efficiency of S-boxes. For an S-box to satisfy the BIC property, all avalanche variables should be independent pair wise for a number of avalanche vectors created by modifying a single bit of plaintext. BIC states that reversing the input bit i modifies the output bits j and k in such a way that the no dependency lies between output bits. This would tend to improve the confusion function's effectiveness. To satisfy the BIC property, the output bits must exhibits independent behavior. Therefore, efforts are being made to decrease the dependency of output bits. The correlation coefficient is used to calculate the degree of independence among avalanche variable pairs. The bit independence of the jth and kth bits of Bei is:

$$BIC(b_j, b_k) = \max_{1 \le i \le n} |corr(b_j^{ei}, b_j^{ei'})| \quad (5)$$

S-box function (h) is defined as: h: {0, 1}n →{0, 1}n .BIC parameter for the S-box function is expressed as follows:

$$BIC(h) = \max_{1 \le j, \; k \le n} BIC(b_j, b_k) \quad (6)$$

Average SAC-BIC score of our optimized S-box is 0.49937. Which is almost optimal and indicates that proposed S-box fulfills the required criteria.

*4.4 Linear Approximation Probability (LP)*

LP is used to evaluate the security of S-box against linear cryptanalysis. S-box provides diffusion and confusion of bits through linear mapping between input and output. Maximum imbalance of an event is determined by LP. The input bit's parity given by the mask γ1 is equal to the output bit's parity given by the mask γ2. Linear approximation probability is represented as:

$$LP = \max_{\gamma1,\gamma2 \neq 0} \left| \frac{\{x \in X | x.\gamma1 = S(x).\gamma2\}}{2^n} - \frac{1}{2} \right| \tag{7}$$

Where the input mask is represented by γ1, γ2 represents the output mask. These masks are used to calculate the linear approximation probability. X denotes the set of all possible inputs and $2^n$ is the total number of elements in the S-Box. An S-box having low linear probability represents high nonlinear mapping and have high resistance against linear cryptanalysis. The maximum optimized S-box LP value is 0.140625 which fulfills the desired criterion.

*4.5 Differential Approximation Probability (DP)*

Differential approximation probability means that output shall have a difference of Δy every time the input is changed by Δx. DP examines the XOR distribution between input and output bit. Variations in output can be obtained from variations in input. For resilience against differential attacks, XOR values of all outputs and inputs must have equal probability. The exclusive-OR distributions among the inputs and outputs of S-box is calculated by:

$$DP\ (\Delta x \rightarrow \Delta y) = \left[\frac{\#\{x \in X\ |\ (S\ (x) \oplus S(x \oplus \Delta x) = \Delta y\}}{2^n}\right] \tag{8}$$

Where X represents the set of all possible input values, $2^n$ is a total number of all the elements in the S-box, Δx and Δy are the input and output differentials. S-boxes with small differentials values are strong and good at resisting differential cryptanalysis. The DP results of Sbox1 are following in the Tab. 4, and we can see that our proposed S-box fulfils the DP criteria.

**Table 4:** DP of the S-box1

| | | | | | | | | | | | | | | | |
|---|---|---|---|---|---|---|---|---|---|---|---|---|---|---|---|
| .000000 | .023438 | .023438 | .023438 | .031250 | .023438 | .023438 | .023438 | .031250 | .023438 | .023438 | .023438 | .023438 | .023438 | .031250 | .023438 |
| .023438 | .023438 | .023438 | .023438 | .023438 | .031250 | .023438 | .031250 | .023438 | .031250 | .023438 | .023438 | .023438 | .023438 | .031250 | .023438 |
| .023438 | .015625 | .031250 | .023438 | .023438 | .039062 | .031250 | .023438 | .023438 | .023438 | .023438 | .023438 | .023438 | .023438 | .023438 | .023438 |
| .023438 | .023438 | .031250 | .031250 | .023438 | .031250 | .023438 | .031250 | .031250 | .031250 | .031250 | .023438 | .031250 | .023438 | .031250 | .046875 |
| .039062 | .023438 | .031250 | .023438 | .023438 | .039062 | .023438 | .023438 | .031250 | .031250 | .023438 | .023438 | .023438 | .023438 | .023438 | .031250 |
| .023438 | .031250 | .023438 | .023438 | .023438 | .023438 | .031250 | .031250 | .023438 | .023438 | .023438 | .031250 | .031250 | .031250 | .023438 | .023438 |
| .023438 | .031250 | .023438 | .023438 | .023438 | .023438 | .023438 | .031250 | .039062 | .023438 | .039062 | .023438 | .031250 | .031250 | .031250 | .023438 |
| .039062 | .023438 | .023438 | .023438 | .023438 | .023438 | .023438 | .023438 | .023438 | .023438 | .023438 | .023438 | .023438 | .023438 | .031250 | .023438 |
| .023438 | .023438 | .039062 | .031250 | .023438 | .023438 | .023438 | .023438 | .023438 | .039062 | .023438 | .023438 | .039062 | .031250 | .023438 | .023438 |
| .023438 | .023438 | .031250 | .031250 | .023438 | .023438 | .023438 | .023438 | .023438 | .023438 | .031250 | .031250 | .023438 | .023438 | .023438 | .031250 |
| .023438 | .023438 | .031250 | .023438 | .023438 | .023438 | .031250 | .023438 | .023438 | .023438 | .031250 | .039062 | .023438 | .031250 | .031250 | .023438 |
| .023438 | .023438 | .023438 | .023438 | .023438 | .031250 | .023438 | .023438 | .023438 | .031250 | .023438 | .023438 | .023438 | .023438 | .031250 | .023438 |
| .031250 | .039062 | .031250 | .023438 | .023438 | .039062 | .031250 | .023438 | .039062 | .031250 | .023438 | .023438 | .031250 | .031250 | .023438 | .031250 |
| .031250 | .023438 | .023438 | .023438 | .023438 | .031250 | .023438 | .023438 | .023438 | .023438 | .023438 | .031250 | .023438 | .023438 | .023438 | .023438 |
| .023438 | .031250 | .031250 | .023438 | .023438 | .031250 | .023438 | .023438 | .023438 | .031250 | .031250 | .023438 | .023438 | .031250 | .031250 | .023438 |
| .023438 | .023438 | .031250 | .023438 | .031250 | .015625 | .039062 | .023438 | .023438 | .031250 | .023438 | .023438 | .039062 | .031250 | .023438 | .023438 |

## *4.6 Substitution Permutation based Image Encryption*

In this section simple Substitution Permutation Network (SPN) used for the image encryption. Detailed steps of the SPN are mentioned bellow. Here for the substitution step optimize S-boxes of the section 3.3 used, we used the P-boxes and keys from the [57].

**Step-1**: Extract the Red, Green and Blue channel values from the color image frames and repeat the following steps-2 to step-4 for every channel(R or G or B).

**Step-2**: Perform the bitwise XOR between the plain pixel values and sub $key_i$ to obtain the key additive pixel values.

**Step-3**: Replace each value of step-2 with entries of the optimized ($Sbox_i$) values.

**Step 4**: Diffuse the each value of step-3 by using ($P\text{-}box_i$) values.

**Step 5**: Repeat step-2 to step-4 for i<16.

**Step 6**: Combine the individual R, G, B encrypted pixel values into a single frame.

The NPCR and UACI are the two frequently used tests of the image cipher to check the strength against various attacks. NPCR, UACI are used to evaluate a large number of plain images to measure the impact of pixel change on the encrypted images. We examined the image encryption results on various standard color images (Lena, pepper, nature, bird, baboon, grapes, sparrow, butterfly). Ideal image encryption algorithm must produce different results when a pixel of the image is slightly varied. NCPR is the rate of change in the number of pixels between two encrypted images obtained from two slightly different images. To achieve maximum sensitivity in an algorithm the value of NCPR should close to 100%. UAIC measure the mean variation of pixel intensity of two encrypted images at same location. Following Tab.5 indicates the values of NPCR and UACI. Constantly our NPCR values are around to 99.63 which is the very good value. Similarly the UACI values are around 33.5 which is also the good value.

**Table 5:** NPCR and UACI results

| Images | Location | NPCR Proposed | UACI Proposed | Images | Location | NPCR Proposed | UACI Proposed |
|---|---|---|---|---|---|---|---|
|  | R | 99.6221 | 33.5514 |  | R | 99.6578 | 33.6534 |
| Lena | G | 99.6127 | 33.5158 | Baboon | G | 99.6256 | 33.6385 |
|  | B | 99.5517 | 33.5212 |  | B | 99.6344 | 33.7265 |
|  | R | 99.6231 | 33.4525 |  | R | 99.6231 | 33.7596 |
| Pepper | G | 99.6462 | 33.4642 | Grapes | G | 99.6652 | 32.7821 |
|  | B | 99.6652 | 33.4935 |  | B | 99.6632 | 33.5063 |
|  | R | 99.5925 | 33.6789 |  | R | 99.6551 | 33.4798 |
| Nature | G | 99.6186 | 33.4987 | Sparrow | G | 99.6225 | 33.4125 |
|  | B | 99.6245 | 33.6506 |  | B | 99.6432 | 32.9098 |
|  | R | 99.6621 | 33.4065 |  | R | 99.6591 | 33.5215 |
| Bird | G | 99.6651 | 32.9154 | Butterfly | G | 99.6652 | 32.9952 |
|  | B | 99.6266 | 32.9365 |  | B | 99.6063 | 33.0563 |

**5 Conclusion**

The IoT-based computationally intelligent schemes are expanding due to the growing concerns of privacy leakage and security attacks in connected systems. It has added a new potential to the internet by enabling communications between objects and humans, making a smarter and more intelligent ecosystem. In this regard, block encryption algorithms have been a standout amongst the most reliable option by which data security is accomplished. The strength of block encryption algorithms against different attacks is dependent on its nonlinear primitive which is called S-boxes. The objective of this research is dynamic generation and optimization of highly secure S-boxes for the block encryption algorithms. For this purpose we used the true random numbers as the entropy source of the proposed method because true random numbers are irreversible, unpredictable, and unreproducible. The proposed method passes all the security evaluation criteria including nonlinearity, linear approximation probability, differential approximation probability, strict avalanche criterion, bit independence criterion, differential analysis, histogram analysis, correlation coefficient tests, NPCR, and UACI tests. The results of our method validate that the proposed dynamic s-boxes are effective for the block encryption algorithms. In the future, we will extend this research to the design of cryptographic key generation technique for the IoT systems.

**Acknowledgment:** We thank the NASA Global Hydrology Resource Center DAAC for providing us the exclusive access on Lightning detection and ranging data DOI: http://dx.doi.org/10.5067/LIS/LDAR/DATA101. Accessed on Dec 04, 2021.

**Funding Statement:** The authors received no specific funding for this study.

**Conflicts of Interest:** The authors declare that they have no conflicts of interest to report regarding the present study.

**Appendix A.**

| Type of Test | P-Value | Conclusion |
|---|---|---|
| 01. Frequency Test (Monobit) | 6.71166621558002 | Non-Random |
| 02. Frequency Test within a Block | 0.21575971288691 | Random |
| 03. Run Test | 0.0 | Non-Random |
| 04. Longest Run of Ones in a Block | 0.16958128602983 | Random |
| 05. Binary Matrix Rank Test | 0.77503219124804 | Random |
| 06. Discrete Fourier Transform (Spectral) Test | 0.44626281343698 | Random |
| 07. Non-Overlapping Template Matching Test | 0.01463534068888 | Random |
| 08. Overlapping Template Matching Test | 0.36023946288540 | Random |
| 09. Maurer's Universal Statistical test | 0.75440628538569 | Random |
| 10. Linear Complexity Test | 0.99963782178561 | Random |
| 11. Serial test: | | |
| | 0.51652456962777 | Random |
| | 0.67625390650010 | Random |
| 12. Approximate Entropy Test | 0.06688206161079 | Random |
| 13. Cummulative Sums (Forward) Test | 3.18952232497357 | Non-Random |
| 14. Cummulative Sums (Reverse) Test | 3.13591282948427 | Non-Random |

15. Random Excursions Test:

| State | Chi Squared | P-Value | Conclusion |
|---|---|---|---|
| -4 | 1.411911703456 | 0.9230058391699 | Random |
| -3 | 2.485400000000 | 0.7786922371169 | Random |
| -2 | 5.354938271604 | 0.3741146665161 | Random |
| -1 | 5.25 | 0.3861379243372 | Random |
| +1 | 2.5 | 0.7764950711233 | Random |
| +2 | 2.567901234567 | 0.7662360828950 | Random |
| +3 | 2.7176 | 0.7434253074909 | Random |
| +4 | 3.424406497292 | 0.6348554712380 | Random |

16. Random Excursions Variant Test:

| State | COUNTS | P-Value | Conclusion |
|---|---|---|---|
| -9.0 | 25 | 0.6995916043003 | Random |
| -8.0 | 36 | 0.3613104285261 | Random |
| -7.0 | 40 | 0.2393165412214 | Random |
| -6.0 | 42 | 0.1658065601940 | Random |
| -5.0 | 30 | 0.4093954862099 | Random |
| -4.0 | 20 | 0.7892680261342 | Random |
| -3.0 | 16 | 1.0 | Random |
| -2.0 | 18 | 0.8382564863858 | Random |
| -1.0 | 16 | 1.0 | Random |
| +1.0 | 18 | 0.7236736098317 | Random |
| +2.0 | 16 | 1.0 | Random |
| +3.0 | 12 | 0.7518296340458 | Random |
| +4.0 | 8 | 0.5929800980174 | Random |
| +5.0 | 5 | 0.5168677495508 | Random |
| +6.0 | 2 | 0.4555450937893 | Random |
| +7.0 | 2 | 0.4924568237728 | Random |
| +8.0 | 1 | 0.4935627897033 | Random |

**Appendix B.**

| 147 | 81  | 50  | 43  | 92  | 245 | 182 | 30  | 237 | 34  | 42  | 39  | 53  | 246 | 219 | 188 |
|-----|-----|-----|-----|-----|-----|-----|-----|-----|-----|-----|-----|-----|-----|-----|-----|
| 1   | 221 | 161 | 76  | 158 | 213 | 160 | 197 | 185 | 164 | 184 | 251 | 223 | 99  | 47  | 27  |
| 172 | 107 | 146 | 83  | 176 | 180 | 171 | 183 | 109 | 56  | 225 | 80  | 64  | 155 | 173 | 63  |
| 15  | 105 | 235 | 163 | 230 | 234 | 193 | 82  | 89  | 177 | 117 | 152 | 216 | 110 | 137 | 91  |
| 162 | 133 | 174 | 122 | 102 | 129 | 108 | 169 | 72  | 116 | 113 | 254 | 22  | 121 | 68  | 204 |
| 249 | 187 | 12  | 115 | 69  | 54  | 95  | 5   | 126 | 94  | 17  | 24  | 77  | 250 | 86  | 154 |
| 222 | 98  | 11  | 165 | 101 | 33  | 87  | 85  | 226 | 25  | 179 | 168 | 201 | 224 | 239 | 70  |
| 31  | 138 | 149 | 96  | 208 | 181 | 106 | 209 | 100 | 215 | 88  | 8   | 139 | 210 | 29  | 104 |
| 40  | 62  | 36  | 52  | 44  | 49  | 220 | 21  | 227 | 218 | 74  | 231 | 18  | 112 | 212 | 150 |
| 206 | 192 | 189 | 211 | 153 | 127 | 10  | 194 | 229 | 6   | 103 | 45  | 134 | 241 | 240 | 217 |
| 214 | 178 | 75  | 142 | 120 | 16  | 252 | 144 | 228 | 55  | 136 | 242 | 118 | 2   | 195 | 71  |
| 202 | 186 | 13  | 28  | 243 | 166 | 148 | 84  | 111 | 141 | 67  | 170 | 167 | 132 | 51  | 123 |
| 93  | 37  | 38  | 248 | 145 | 79  | 135 | 175 | 32  | 78  | 156 | 238 | 203 | 236 | 61  | 232 |
| 23  | 130 | 143 | 3   | 90  | 35  | 233 | 46  | 196 | 151 | 48  | 159 | 19  | 20  | 119 | 65  |
| 73  | 14  | 131 | 200 | 97  | 190 | 128 | 157 | 244 | 205 | 26  | 207 | 125 | 199 | 41  | 140 |
| 58  | 57  | 4   | 66  | 253 | 59  | 247 | 60  | 114 | 9   | 198 | 255 | 124 | 191 | 7   | 0   |

**Appendix C.**

**Algorithm: ReverseSbox (S-box)**

**in:** 2D array of integers, sbox[16][16];
**out:** 2D array of integers, ReverseSbox [16][16];
1: ReverseSbox → |16||16|
2: **for** row → 0 … (16) **do**
3:     **for** col → 0 … (16) **do**
4:         rowIS → sbox $_{row,col}$ **div** 16
5:         colIS → sbox $_{row,col}$ **mod** 16
6:         value → row*16+col
7:             ReverseSbox $_{rowIS,colIS}$ →  value
8:     **end for**
9: **end for**
10: **return** ReverseSbox